\documentclass[10pt,abstract]{sensys-proc}

\usepackage{textcomp}
\usepackage[compact]{titlesec}
\usepackage[font=small,labelfont=bf]{caption}
\usepackage{graphicx}

\begin{document}

\title{Demo abstract: NILMTK v0.2: A Non-intrusive Load Monitoring Toolkit for Large Scale Data Sets}
%
%
%
%
%

\numberofauthors{1} 
%
\author{Jack Kelly$^1$, Nipun Batra$^2$, Oliver Parson$^3$, Haimonti Dutta$^4$, William Knottenbelt$^1$,\\ Alex Rogers$^3$, Amarjeet Singh$^2$, Mani Srivastava$^5$\\ \\
\small $^1$ Imperial College London ~\{jack.kelly,~w.knottenbelt\}@imperial.ac.uk\\
\small $^2$Indraprastha Institute of Information Technology Delhi, India ~\{nipunb,~amarjeet\}@iiitd.ac.in\\
\small $^3$ University of Southampton ~\{osp,~acr\}@ecs.soton.ac.uk\\
\small $^4$ CCLS Columbia ~haimonti@ccls.columbia.edu\\
\small $^5$ UCLA ~mbs@ucla.edu\\
}

\crdata{978-1-4503-3144-9}
\doidata{10.1145/2674061.2675024}
\conferenceinfo{BuildSys'14,} {November 5--6, 2014, Memphis, TN, USA.}
\CopyrightYear{2014}


\maketitle

\begin{abstract}
In this demonstration, we present an open source toolkit for evaluating non-intrusive load monitoring research; a field which aims to disaggregate a household's total electricity consumption into individual appliances. The toolkit contains: a number of importers for existing public data sets, a set of preprocessing and statistics functions, a benchmark disaggregation algorithm and a set of metrics to evaluate the performance of such algorithms. Specifically, this release of the toolkit has been designed to enable the use of large data sets by only loading individual chunks of the whole data set into memory at once for processing, before combining the results of each chunk.
\end{abstract}



\section{Introduction}

The field of non-intrusive load monitoring (NILM) was founded by Hart over 30 years ago \cite{hart_1992}. In recent years, the field has rapidly expanded due to smart meter deployments across many countries and release of multiple data sets for NILM evaluation. However, new approaches have typically preprocessed the existing data sets in different ways, been compared with different benchmark algorithms, and evaluated using different accuracy metrics. Consequently, it was not possible to quantitatively compare the accuracy of any two state-of-the-art disaggregation algorithms.

In April 2014, an open source toolkit for non-intrusive load monitoring (NILMTK v0.1) was released to overcome these limitations~\cite{nilmtk}. The toolkit contained a number of importers for existing public data sets, a set of preprocessing and statistics functions, two benchmark disaggregation algorithms and a set of metrics to evaluate the performance of such algorithms. However, the toolkit was designed to handle the relatively small data sets (less than 10 households) which were available at the time of release. As such, the toolkit was not suitable for use with larger data sets (hundreds of households) which have been released since (e.g.\ WikiEnergy data set). As a result, it has not been possible to evaluate energy disaggregation approaches at a sufficient scale so as to investigate the extent of their generality.

To address this shortcoming, we present a new release of the toolkit (NILMTK v0.2) which is able to evaluate energy disaggregation algorithms using arbitrarily large data sets.  Rather than loading the entire data set into memory (as in v0.1), the aggregate data is loaded in chunks and the output of the disaggregation algorithm is saved to disk chunk-by-chunk. As a result, we are able to demonstrate data set statistics and disaggregation for the WikiEnergy data set, which contains 239 households of aggregate and individual appliance power data. In addition to scalability improvements, v0.2 also includes support for a rich data set metadata description format, as well as a number of usability improvements and many software design improvements.

\section{NILMTK v0.2}

NILMTK~v0.2 improves on NILMTK~v0.1~\cite{nilmtk} in several important ways.  Indeed, these changes are so fundamental that NILMTK~v0.2 is a complete re-write of NILMTK. We now present the changes in NILMTK~v0.2.

\subsection{Scalability}
\begin{figure}
\vspace{-6pt}
\centering 
\includegraphics[scale=0.7]{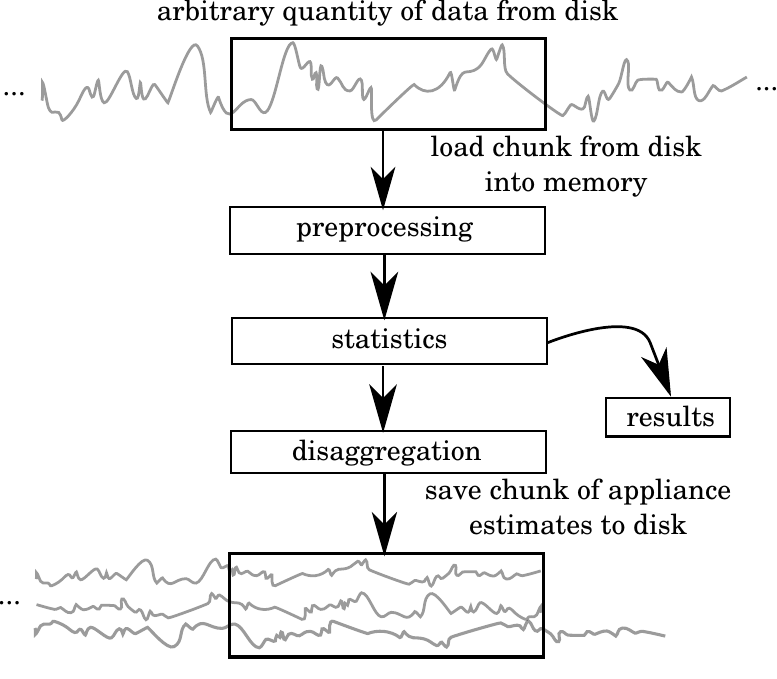}
\caption{\small{NILMTK v0.2 can process an arbitrary quantity of data by loading data from disk in chunks.  This figure illustrates the loading of a chunk of aggregate data from disk (top) and then pushing this chunk through a processing pipeline which ends in saving appliance estimates to disk chunk-by-chunk.}}
\label{fig:pipeline}
\end{figure}

The most important change is that v0.2 can handle data sets of arbitrary size whilst v0.1 can only handle data sets which fit into system memory.  v0.2 achieves this `out-of-core' functionality by lazily loading data in chunks (Fig.~\ref{fig:pipeline}).  When a user first opens a data set, only the metadata is loaded into memory.  The user then builds a processing pipeline (e.g.\ preprocess the data, then calculate some statistics, then train a disaggregation algorithm).  NILMTK loads data from disk into the pipeline in chunks and only holds one chunk in memory at any one time.  Under the hood, we have two families of statistics classes: \texttt{Node} classes perform calculations on each chunk (e.g.\ calculating the total energy) and \texttt{Results} classes know how to merge results from multiple chunks (e.g.\ sum the total energy calculated for each chunk).  NILMTK supports both batch and on-line algorithms and, if required, the subsequent data chunk can be `peeked' into by passing the argument \texttt{n\_look\_ahead\_rows} to \texttt{DataStore.load()}.

During disaggregation, the aggregate data is loaded in chunks and the estimated power demand for each appliance is saved to disk chunk-by-chunk so disaggregation can be performed on arbitrarily large time series. This is in complete contrast to NILMTK v0.1, which eagerly loads the entire aggregate time series into memory as well as storing all appliance estimates in memory.  Saving disaggregation estimates to disk also lays the foundation for sharing raw disaggregation output to allow researchers to do more rigorous comparisons of different disaggregation approaches.  The NILMTK metrics functions load the appliance estimates from disk, allowing them to work on arbitrarily large data sets and also allowing users to take advantage of the metrics functions without using any other part of the NILMTK pipeline if they wish.

\subsection{Rich metadata support}
In addition to the out-of-core functionality, NILMTK~v0.2 also offers much richer metadata support than v0.1 by integrating with the NILM Metadata project~\cite{NILM_Metadata}.  For example, NILMTK can now handle electricity meter wiring hierarchies of arbitrary complexity, knows the mapping between appliance types and appliance categories (e.g.\ a `fridge' is a `cold' appliance), and uses a controlled vocabulary for appliance type names.

\subsection{Improved design and usability}
With NILMTK~v0.2, we aimed to make better use of object oriented programming to hide complexity from the user and compartmentalise functionality.  As an example of hiding complexity: sometimes it is useful to group multiple electricity meters together and treat them as a single meter.  For example, north-American homes are fed with a single phase power supply split into two 120 volt supplies.  Hence each north-American home tends to have two mains meters.  It is often useful to sum these two meters together.  NILMTK v0.2 provides an \texttt{ElecMeter} class for representing single meters and a \texttt{MeterGroup} class for representing multiple \texttt{ElecMeters}.  These two classes share many of the same methods so they can be used interchangeably in many contexts.  The \texttt{MeterGroup} class also allows for the sophisticated selection of meters using any metadata property.  For example, all \texttt{cold} appliances could be selected and their total energy calculated with \\ \texttt{elec.select(category=\textquotesingle lighting\textquotesingle).total\_energy()}.

\subsection{Improved warnings}
An experimental feature of NILMTK~v0.2 is a framework for performing precondition checks for each statistics function.  The motivation is that some statistics functions require the data to be preprocessed in a certain way.  In NILMTK~v0.1, if a user attempts to run a statistics function on incorrectly preprocessed data then no warning is raised and the wrong answer will be produced.  In v0.2, statistics classes declare a set of preconditions.  These preconditions are checked against the data set metadata and the previous nodes in the processing pipeline.  If any preconditions are not met then the user is warned immediately (before any time series data has been loaded from disk).



\subsection{Data set importers}
NILMTK v0.2 includes data set converts for REDD~\cite{redd}, iAWE~\cite{iawe}, WikiEnergy (\texttt{wiki-energy.org}) and GreenD~\cite{greend}, COMBED~\cite{combed} and UK-DALE~\cite{ukdale}.  The GreenD importer was written by the data set authors and was the first significant code contribution from outside of the core NILMTK contributors.

\section{Conclusions}
In this demonstration, we presented an open source toolkit for evaluating non-intrusive load monitoring research using large scale data sets. v0.1 of the toolkit was released in April 2014 and v0.2 was released in July 2014.


%
\bibliographystyle{abbrv}
\begin{small}
\bibliography{sigproc}  
\end{small}
%
%



\newif\ifcameraready
\camerareadytrue 

\ifcameraready

\else 
\clearpage

\section{Script}

In this section we describe our plan for demonstrating NILMTK v0.2, which follows a similar format to our demonstration of NILMTK~v0.1\footnote{\url{http://goo.gl/AyA2ct}} at the 2014 NILM workshop.\footnote{\url{http://nilmworkshop.org/}} Our demonstration will be based on an IPython notebook,\footnote{\url{http://ipython.org/notebook.html}} which allow code and output to be embedded in a single document and thus are ideal for demonstrating reproducible research and open science. Our demonstration will highlight the following six new contributions (discussed in Section 2) made to NILMTK~v0.2: i) scalability ii) rich metadata support iii) improved design and usability iv) improved warnings and v) data set importers. To illustrate these contributions, our demonstration IPython notebook will follow the following steps:
\begin{enumerate}
\item Convert an existing data set such as REDD to NILMTK data format.
\item Load WikiEnergy data set (which is too large fit into memory in its entirety) into NILMTK.
\item Diagnose and mask out missing data in a subset of buildings from WikiEnergy data set.
\item Compute statistics pertinent to NILM, such as percentage energy sub-metered, entropy, simultaneous appliance switches, correlation among different appliances, among others.
\item Compute aggregation statistics, such as category-wise energy consumption (hot vs cold), hourly median power usage, among others.
\item Divide the data into train and test sets.
\item Using the benchmark combinatorial optimisation algorithm, train on the train set and disaggregate on the test set.
\item Export the model and disaggregation results for reuse by other researchers.
\item Compute disaggregation metrics to evaluate the disaggregation result across multiple buildings.
\end{enumerate}

\section{Logistical requirements}
Since our demo would be software based, logistical requirements for the demonstration are modest. A poster board, a small table space (1m x 1m), a large computer display (although the demo can be performed from a laptop if this is not available), and a power connection are the only requirements.
\fi

\end{document}